\documentclass[a4paper]{jpconf}
\usepackage{graphicx}
\usepackage{amsmath,amssymb,amscd}
\usepackage{multirow}

\begin{document}
\title{Discrete Glimpses \\ {\small of the Physics Landscape after the Higgs Discovery}}

\author{John Ellis}

\address{Physics Department, King's College London, Strand, London WC2R 2LS, U.K;
Theory Division, Physics Department, CERN, CH 1211 Geneva 23, Switzerland}

\ead{John.Ellis@cern.ch}

\begin{abstract}
What is the Higgs boson telling us? What else is there? How do we find it?
This talk discusses these current topics in particle physics in the wake of the Higgs discovery,
with particular emphasis on the discrete symmetries CP and R-parity, not forgetting
flavour physics and dark matter, and finishing with some remarks about possible
future colliders. \\
~~\\
{\tt KCL-PH-TH/2015-02, LCTS/2015-01, CERN-PH-TH/2015-008}
\end{abstract}

\section{The Story so far}

\begin{figure}[htb]
\centering
\includegraphics[height=3in]{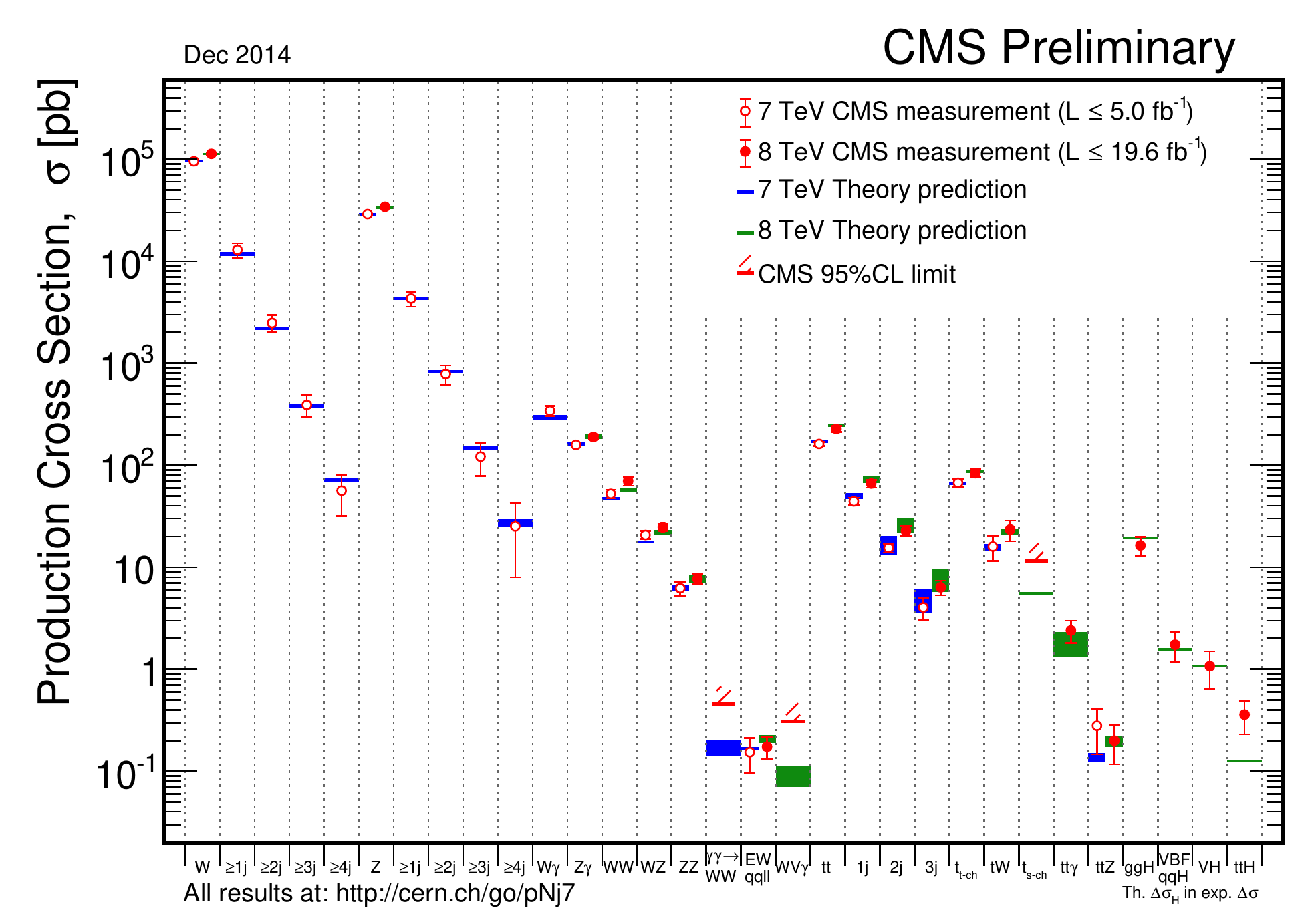}
\caption{\it A compilation of cross sections at the LHC measured by the CMS Collaboration~\protect\cite{CMSxs}.}
\label{fig:sections}
\end{figure}

Run 1 of the LHC has brought the apotheosis of the Standard Model (SM). The SM has predicted 
successfully many cross sections for particle and jet production measured at the LHC~\cite{CMSxs}, 
as seen in Fig.~\ref{fig:sections}. These successes include QCD jet production cross sections, which agree with
SM predictions over large ranges in energy and many orders of magnitude,
measurements of single and multiple $W^\pm$ and $Z^0$ production, as well as
multiple measurements of top quark production, both pairwise and singly and in association with
vector bosons. Moreover, Run 1 of the LHC has also brought the discovery by CMS and ATLAS of a (the?) Higgs boson~\cite{Higgs},
whose production has by
now been observed in three different production channels, as also seen in Fig.~\ref{fig:sections}, again with cross sections
in agreement with the SM predictions. A major theme of this talk is what we already know about this newly-discovered particle.

%%%%%%%%%%%%%%%%%%%%%%%%%%%%%%%%%%%%%%%%%%%%%%%%%%%%%%%%%%%%%%%%%%%%%%%%%
%%
%%   use this format to include an .eps figure into your paper
%%
%%%%%%%%%%%%%%%%%%%%%%%%%%%%%%%%%%%%%%%%%%%%%%%%%%%%%%%%%%%%%%%%%%%%%%%%%%%

\begin{figure}[htb]
\centering
\includegraphics[height=2.3in]{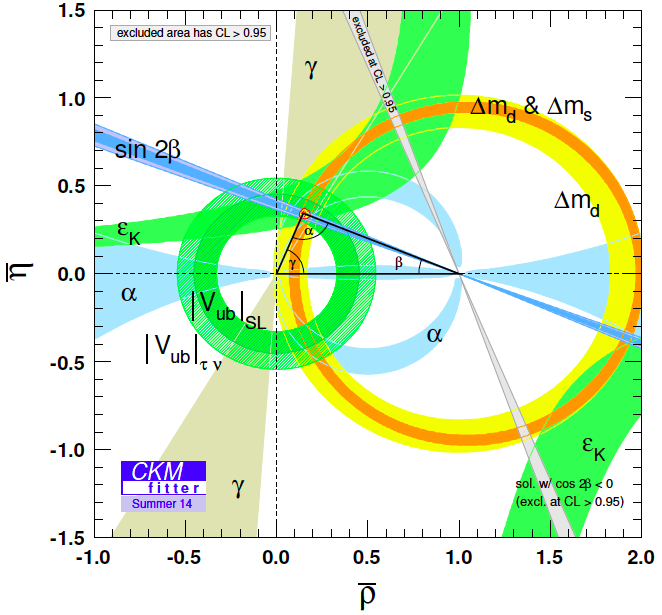}
\includegraphics[height=2.3in]{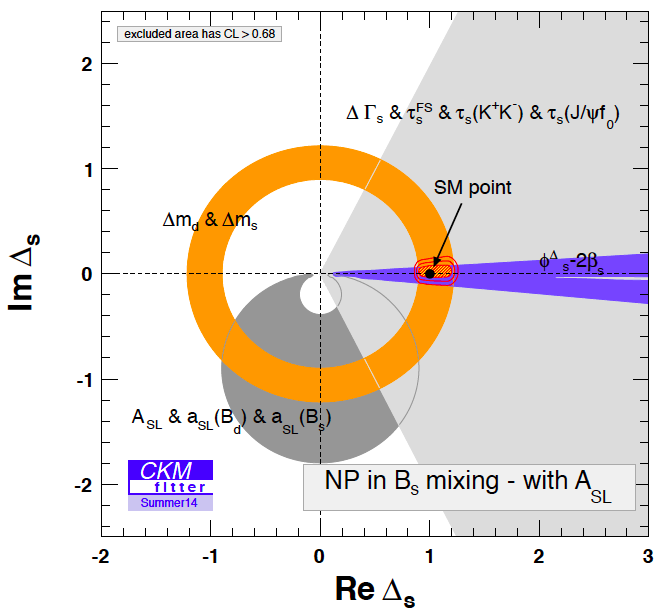}
\caption{\it Left panel: Flavour and CP violation measurements generally agree well with the CKM paradigm.
Right panel: Experimental constraint on a possible non-Standard Model contribution to $B_s$ mixing~\protect\cite{CKMFitter}.}
\label{fig:hs}
\end{figure}

\begin{figure}[htb]
\centering
\includegraphics[height=2.3in]{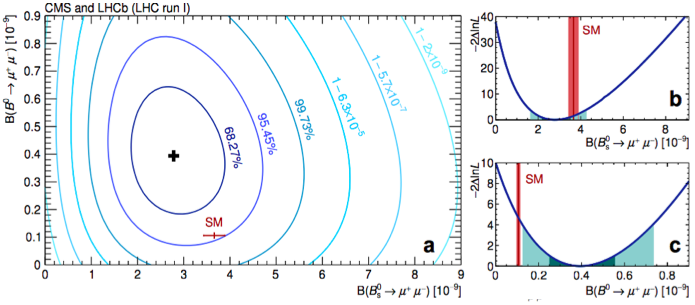}
\caption{\it Panel {\bf a}: measurements by the CMS and LHCb Collaborations~\protect\cite{Bsmumu} 
of $B_{s,d} \to \mu^+ \mu^-$ decays. Panel {\bf b}: The CMS and LHCb Collaborations see a
clear signal for $B_s \to \mu^+ \mu^-$ decay. Panel {\bf c}: They also see
a possible hint of $B_d \to \mu^+ \mu^-$ decay.}
\label{fig:Bmumu}
\end{figure}

Meanwhile, over there in flavour space,
the Cabibbo-Kobayashi-Maskawa (CKM) description of flavour mixing and CP violation
is also a pillar of the SM, but here the picture is more complex. It is in general very successful,
as seen in the left panel of Fig.~\ref{fig:hs}~\cite{CKMFitter},
but there are a number of anomalies.
On the one hand, the SM predicted successfully the branching ratio for the rare decay $B_s \to \mu^+ \mu^-$:
\begin{equation}
BR(B_s \to \mu^+ \mu^-) \; = \; 2.8^{+0.7}_{-0.6} \times 10^{-9} \, ,
\label{Bsmumu}
\end{equation}
measured by the CMS and LHCb Collaborations~\cite{Bsmumu}: see Fig.~\ref{fig:Bmumu}.
However, the joint CMS and LHCb analysis~\cite{Bsmumu} also has an suggestion of a $B_d \to \mu^+ \mu^-$ signal
that is considerably larger than the SM prediction (which is an ironclad prediction also of models with minimal flavour violation (MFV),
 including many SUSY scenarios):
\begin{equation}
BR(B_d \to \mu^+ \mu^-) \; = \; 3.9^{+1.6}_{-1.4} \times 10^{-10} \, ,
\label{Bdmumu}
\end{equation}
as also seen in Fig.~\ref{fig:Bmumu}. This will be something to watch during Run~2:
Parhaps MFV is not the whole story? {\it Wait and see!}

%%%%%%%%%%%%%%%%%%%%%%%%%%%%%%%%%%%%%%%%%%%%%%%%%%%%%%%%%%%%%%%%%%%%%%%%%
%%
%%   use this format to include an .eps figure into your paper
%%
%%%%%%%%%%%%%%%%%%%%%%%%%%%%%%%%%%%%%%%%%%%%%%%%%%%%%%%%%%%%%%%%%%%%%%%%%%%

%%%%%%%%%%%%%%%%%%%%%%%%%%%%%%%%%%%%%%%%%%%%%%%%%%%%%%%%%%%%%%%%%%%%%%%%%
%%
%%   use this format to include an .eps figure into your paper
%%
%%%%%%%%%%%%%%%%%%%%%%%%%%%%%%%%%%%%%%%%%%%%%%%%%%%%%%%%%%%%%%%%%%%%%%%%%%%

There is scope elsewhere for deviations from CKM predictions:
for example, the data allow an important contribution to the mixing amplitude
for $B_s$ mesons from physics beyond the SM (BSM),
as seen in the right panel of Fig.~\ref{fig:hs}~\cite{CKMFitter}.
Moreover, there are several indications of anomalies in flavour physics:
for example, the branching ratio for $B^\pm \to \tau^\pm \nu$ decay differs from the SM prediction
by $\sim 2 \sigma$, and there are issues with $e - \mu$ universality in semileptonic $B$ decays~\cite{nonuniv}.
The most significant anomaly appears in the $P_5^\prime$ angular distribution for
$B^0 \to K^{*0} \mu^+ \mu^-$~\cite{P5prime}, though the non-perturbative corrections need to be understood
better. It has recently been suggested that these may be related to the intriguing excess in $H \to \mu \tau$
decay reported by the CMS Collaboration~\cite{CMSHmutau}, as discussed later.
Also worth noting are discrepancies in the determinations of the
$V_{ub}$ CKM matrix element, and there is still an anomaly in the
diimuon asymmetry at the Tevatron~\cite{dimuon}.
On the other hand, some anomalies do seem to be going away, such as the forward-backward asymmetry in
$t \bar{t}$ production, which now agrees with higher-order QCD calculations~\cite{ttbarAFB}, as does the
 $t \bar{t}$ rapidity asymmetry measured at the LHC. However, the discrete charm of 
flavour physics leaves many issues to be addressed during LHC Run~2 and at SuperKEK-B.

One of the principal focuses during Run~2 of the LHCwill be the
more detailed study of the Higgs boson and probes whether its properties deviate from
SM predictions, e.g., in the flavour sector. As I discuss later, the
measurement of the Higgs mass has produced new reasons to expect BSM physics, and the search
for BSM physics will start anew at Run~2, with its greatly increased centre-of-mass energy and
increased integrated luminosity. My personal favourite candidate for BSM physics is supersymmetry
(SUSY), and I also discuss later in this talk how SUSY models are constrained by flavour physics, as well as by the
observations to date of the Higgs boson and searches for BSM physics with Run-1 data.
This talk concludes with some remarks about searches for particle dark matter at the LHC and
elsewhere, and some advertisements for possible future colliders.

\section{Higgs Physics}

\subsection{Mass Measurements}

The mass of the Higgs boson can be measured most accurately
in the $\gamma \gamma$ and $Z Z^* \to 2 \ell^+ 2 \ell^-$ final states, and ATLAS and CMS both
report accurate measurements in both these final states. ATLAS measures~\cite{ATLASmH}
\begin{eqnarray}
%H \to \gamma \gamma: m_H & = \; 125.98 \pm 0.42 \pm 0.28~{\rm GeV} = \; 125.98 \pm 0.50~{\rm GeV}\, , \nonumber \\
%H \to Z Z^*: m_H & = \; 125.51 \pm 0.52 \pm 0.04~{\rm GeV} = \; 125.51 \pm 0.52~{\rm GeV}\, , \nonumber \\
{\rm ATLAS~combined:} ~m_H & = \; 125.36 \pm 0.37 \pm 0.18~{\rm GeV} \; = \; 125.36 \pm 0.41~{\rm GeV}\, ,
\label{ATLASm}
\end{eqnarray}
and CMS measures~\cite{CMSmH}
\begin{eqnarray}
%H \to \gamma \gamma: m_H & = \; 124.70 \pm 0.31 \pm 0.15~{\rm GeV} \; = \; 124.70^{+0.35}_{-0.34}~{\rm GeV}\, , \nonumber \\
%H \to Z Z^*: m_H & = \; 125.6 \pm 0.4 \pm 0.2~{\rm GeV}\, = \; 125.6 \pm 0.4~{\rm GeV} \, , \nonumber \\
{\rm CMS~combined:} ~m_H & = \; 125.03^{+ 0.26}_{-0.27}~^{+ 0.13}_{- 0.15}~{\rm GeV} \; = \; 125.03 \pm 0.30~{\rm GeV}\, .
\label{CMSm}
\end{eqnarray}
Some interest has been generated by the differences in the masses measured in these channels,
but these have opposite signs in the two experiments:
\begin{eqnarray}
{\rm ATLAS}: \; \Delta m_H & = \; 1.47 \pm 0.67 \pm 0.18~{\rm GeV} \, , \nonumber \\
{\rm CMS}: \; \Delta m_H & = \; - 0.9 \pm 0.4 \pm 0.2 ^{+0.34}_{-0.35}~{\rm GeV} \, ,
\label{Deltam}
\end{eqnarray}
so are presumably statistical and/or systematic artefacts.
Combining naively the ATLAS and CMS measurements yields
\begin{equation}
m_H \; = \; 125.15 \pm 0.24~{\rm GeV}.
\label{mH}
\end{equation}
In addition to being a fundamental measurement in its own right, and casting light on the possible validity of various BSM models,
the precise value of $m_H$ is also important for the stability of the electroweak vacuum in the Standard Model,
as discussed later.

\subsection{The Higgs Spin and Parity}

\begin{figure}[htb]
\centering
\includegraphics[height=2.3in]{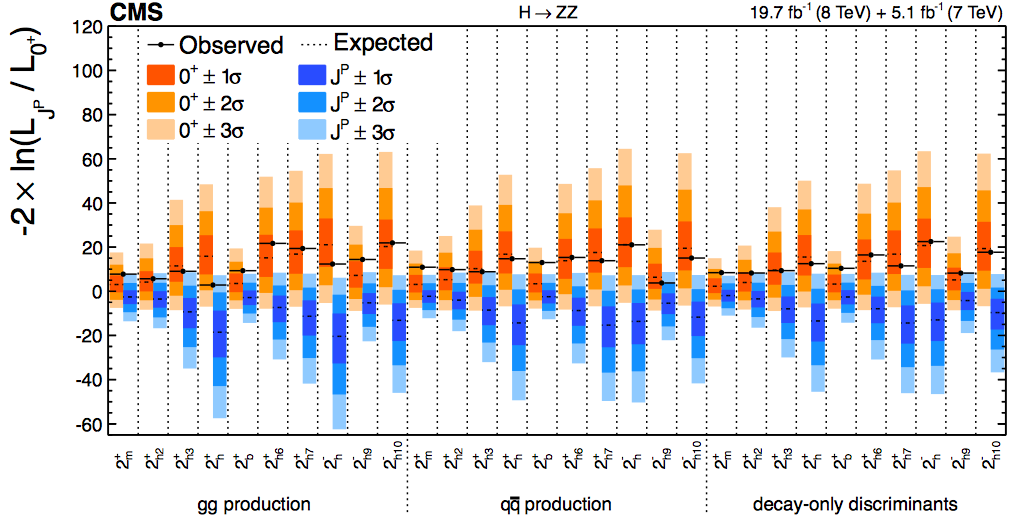}
\caption{\it Tests of spin-parity hypotheses for the Higgs boson. All favour strongly the $0^+$ assignment
expected in the SM~\protect\cite{CMSHspin}.}
\label{fig:Higgsspin}
\end{figure}

The fact that the Higgs boson has been observed to decay into $\gamma \gamma$ excludes spin 1,
and spin 0 is expected, but spins 2 and higher are also possible in principle. There have been many
probes of the Higgs spin~\cite{ATLASHspin,CMSHspin,CDFHspin},
including its production and decay rates~\cite{ESYrates}, the kinematics of its production
in association with massive vector bosons~\cite{EHSY}, as well as angular distributions in its decays into $W^+ W^-$,
$ZZ$ and $\gamma \gamma$~\cite{EFHSY}. The results of many of these tests are shown in Fig.~\ref{fig:Higgsspin}.
By now there is overwhelming evidence that the Higgs boson has spin 0 and that its couplings to
$W^+ W^-$ and $ZZ$ are predominantly CP-even.

%%%%%%%%%%%%%%%%%%%%%%%%%%%%%%%%%%%%%%%%%%%%%%%%%%%%%%%%%%%%%%%%%%%%%%%%%
%%
%%   use this format to include an .eps figure into your paper
%%
%%%%%%%%%%%%%%%%%%%%%%%%%%%%%%%%%%%%%%%%%%%%%%%%%%%%%%%%%%%%%%%%%%%%%%%%%%%

\begin{figure}[htb]
\centering
\includegraphics[height=2.29in]{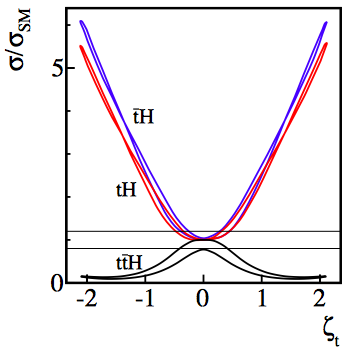}
\includegraphics[height=2.31in]{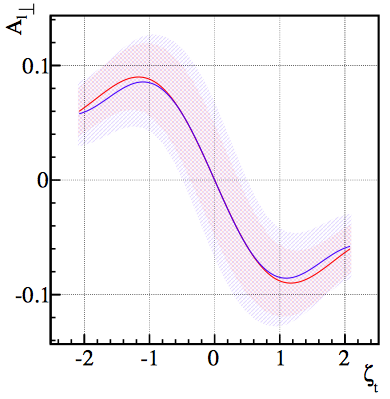}
\caption{\it Left panel: The effects of a CP-violating coupling on the $H t {\bar t}$, $H t$ and $H {\bar t}$ production
cross sections. Right panel: A CP-violating transverse polarization asymmetry observable in 
$H t$ and $H {\bar t}$ production~\protect\cite{EHST}.}
\label{fig:tHCP}
\end{figure}

On the other hand, it is possible that there may be an admixture of CP-odd couplings,
and their fraction may depend on the particle whose coupling to the Higgs boson are
being probed. In particular, the leading CP-odd $H$ coupling to fermions would have the
{\it same} (zero) dimension as the leading CP-even coupling, whereas the leading CP-odd $H$ coupling to
massive vector bosons would have {\it higher} dimension than the leading CP-even coupling,
so it may be more suppressed. Various ways to probe CP violation in the $H \tau^+ \tau^-$ couplings
have been proposed~\cite{Askewetal}, and it is also possible to probe CP violation in the $H t {\bar t}$
couplings~\cite{EHST}. These would affect the total cross sections for associated $H t {\bar t}$,
$H t$ and $H {\bar t}$ production, as seen in the left panel of Fig.~\ref{fig:tHCP} as
functions of $\zeta_t \equiv$ arctan(CP-odd coupling/CP-even coupling). If $\zeta_t \ne 0$,
a CP-violating transverse polarization asymmetry is in principle observable in $H t$ and $H {\bar t}$ production,
as shown in the right panel of Fig.~\ref{fig:tHCP}.

%%%%%%%%%%%%%%%%%%%%%%%%%%%%%%%%%%%%%%%%%%%%%%%%%%%%%%%%%%%%%%%%%%%%%%%%%
%%
%%   use this format to include an .eps figure into your paper
%%
%%%%%%%%%%%%%%%%%%%%%%%%%%%%%%%%%%%%%%%%%%%%%%%%%%%%%%%%%%%%%%%%%%%%%%%%%%%

\subsection{Higgs Couplings}

\begin{figure}[htb]
\centering
\includegraphics[height=2.3in]{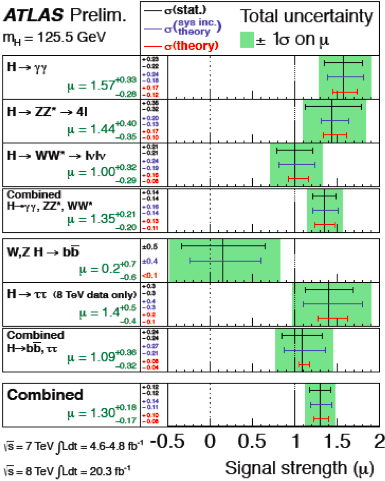}
\includegraphics[height=2.3in]{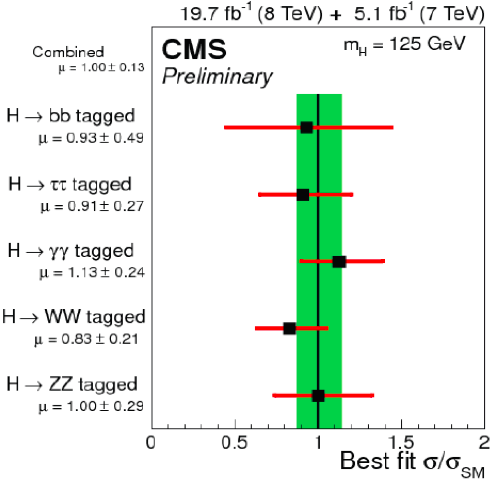}
\caption{\it The Higgs signal strengths $\mu$, normalised to unity for the SM, as measured by
ATLAS~\protect\cite{ATLASmu} (left panel) and CMS~\protect\cite{CMSmH} (right panel).}
\label{fig:strengths}
\end{figure}

As seen in Fig.~\ref{fig:strengths}, the strengths of the Higgs signals measured
by ATLAS and CMS individual channels are generally
compatible with the SM predictions within the statistical fluctuations~\cite{ATLASmu,CMSmu,CMSmH},
which are inevitably large at this stage. Combining their measurements
in the $\gamma \gamma$,
$Z Z^*$, $W W^*$, $b \bar{b}$ and $\tau^+ \tau^-$ channels, ATLAS
and CMS report the following overall signal strengths:
\begin{eqnarray}
{\rm ATLAS:} ~\mu & = \; 1.30 \pm 0.12 \pm 0.10 \pm 0.09 \, , \nonumber \\
{\rm CMS:} ~ \mu & = \; 1.00 \pm 0.09~^{+ 0.08}_{- 0.07} \pm 0.07 \, .
\label{mu}
\end{eqnarray}
These averages are again quite compatible with each other and with the SM,
and measurements at the Tevatron are also compatible with SM predictions for the Higgs boson~\cite{TevatronH}.

%%%%%%%%%%%%%%%%%%%%%%%%%%%%%%%%%%%%%%%%%%%%%%%%%%%%%%%%%%%%%%%%%%%%%%%%%
%%
%%   use this format to include an .eps figure into your paper
%%
%%%%%%%%%%%%%%%%%%%%%%%%%%%%%%%%%%%%%%%%%%%%%%%%%%%%%%%%%%%%%%%%%%%%%%%%%%%

\begin{figure}[htb]
\centering
\includegraphics[height=2.15in]{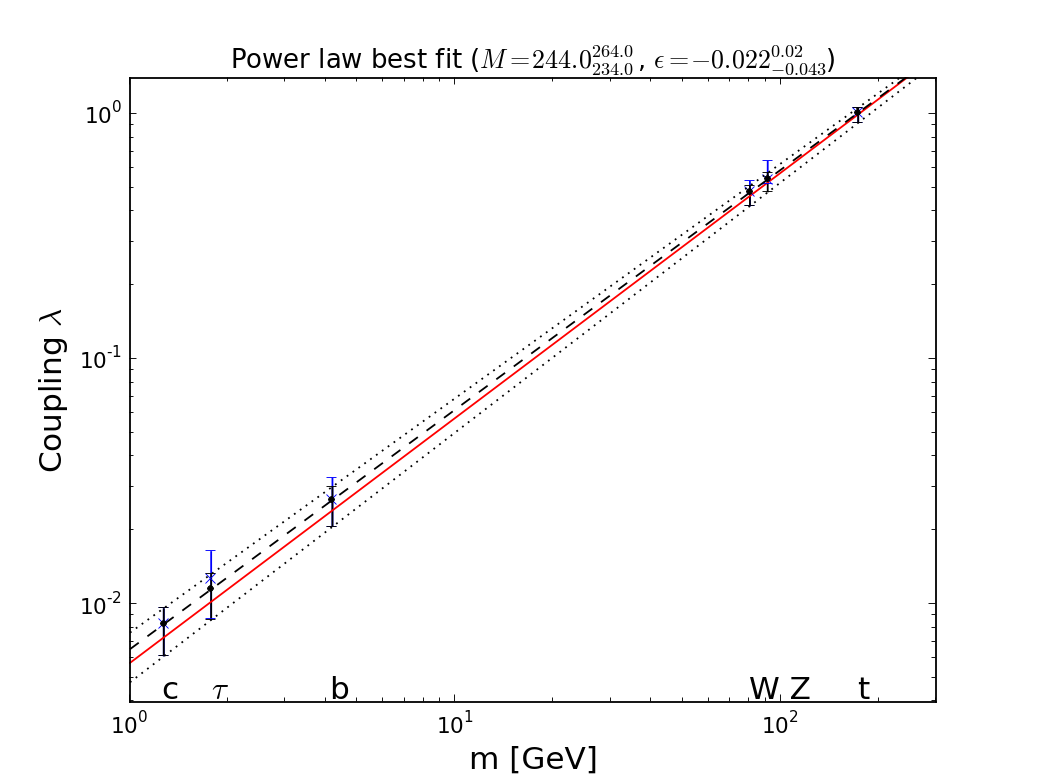}
\caption{\it A global fit to the $H$ couplings of the form (\protect\ref{epsilon}) (central values as dashed and
$\pm$1$\sigma$ values as dotted lines), which is very compatible
with the expected linear mass dependence for fermions and quadratic mass dependence for bosons (solid
red line)~\protect\cite{EY3}.}
\label{fig:Mass_dependence}
\end{figure}

One distinctive feature of the Higgs couplings to other particles in the SM is that they should be related to their masses:
linearly for fermions, quadratically for bosons, and be proportional to the Higgs vev $v = 246$~GeV.
These predictions are verified indirectly by the
measurements in Fig.~\ref{fig:strengths}, but one may also test these predictions directly, as seen
in Fig.~\ref{fig:Mass_dependence}. This shows the result of a global fit to the data
parametrising the Higgs couplings as~\cite{EY3}
\begin{equation}
\lambda_f \; = \; \sqrt{2} \left( \frac{m_f}{M} \right)^{(1 + \epsilon)}, \; \; 
g_V \; = \; 2 \left( \frac{M_V^{2(1 + \epsilon)}}{M^{(1 + \epsilon)}} \right) \, .
\label{epsilon}
\end{equation}
As seen in the left panel of Fig.~\ref{fig:Mass_dependence}, the data yielded
\begin{equation}
\epsilon \; = \; - 0.022^{+0.020}_{-0.043}, \; \; M \; = \; 244^{+20}_{-10}~{\rm GeV}, \,
\label{epsilonM}
\end{equation}
quite compatible with the SM predictions $\epsilon = 0$, $M = 246$~GeV.
Similar results have also been found recently in an analysis by the CMS Collaboration~\cite{CMSmu}.
It seems that Higgs couplings have a very similar flavour structure
to particle masses.

%%%%%%%%%%%%%%%%%%%%%%%%%%%%%%%%%%%%%%%%%%%%%%%%%%%%%%%%%%%%%%%%%%%%%%%%%
%%
%%   use this format to include an .eps figure into your paper
%%
%%%%%%%%%%%%%%%%%%%%%%%%%%%%%%%%%%%%%%%%%%%%%%%%%%%%%%%%%%%%%%%%%%%%%%%%%%%

A related aspect of the SM is the expectation is that flavour should be conserved to a very good
approximation in Higgs couplings to fermions. This is indeed consistent with the upper limits on low-energy effective
flavour-changing interactions, but these would allow also lepton-flavour-violating
Higgs couplings much larger than the SM predictions, so looking for such
interactions is a possible window on BSM physics. We estimated on the basis of low-energy data that
the branching ratios for $H \to \mu \tau$ and $H \to e \tau$ decays could each be
as large as ${\cal O}(10)$\%, i.e., as large as BR$(H \to \tau \tau$, whereas the branching ratio for $H \to \mu e$
could only be $\lesssim 10^{-5}$~\cite{BEI}. The CMS Collaboration has recently found~\cite{CMSHmutau}
\begin{equation}
{\rm BR}(H \to \mu \tau) \; = \; 0.89^{+0.40}_{-0.37} \, \% \, ,
\label{Hmutau}
\end{equation}
corresponding to a background-only $p$-value of 0.007, a $\sim$2.46$\sigma$ effect.
LHC Higgs measurements are therefore already testing SM
flavour physics predictions more stringently than previous low-energy experiments, and we are on tenterhooks
to see corresponding results from ATLAS and from Run~2 of the LHC!

%%%%%%%%%%%%%%%%%%%%%%%%%%%%%%%%%%%%%%%%%%%%%%%%%%%%%%%%%%%%%%%%%%%%%%%%%
%%
%%   use this format to include an .eps figure into your paper
%%
\begin{figure}[htb]
\centering
\includegraphics[height=2.3in]{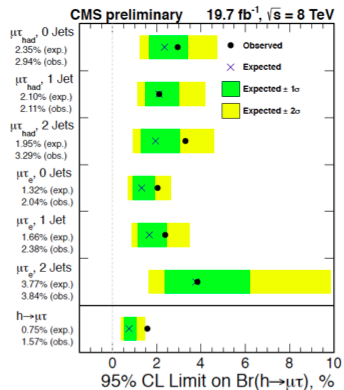}
\caption{\it Results from the CMS search for $H \to \mu \tau$ decay~\protect\cite{CMSHmutau}.}
\label{fig:Hmutau}
\end{figure}
%%%%%%%%%%%%%%%%%%%%%%%%%%%%%%%%%%%%%%%%%%%%%%%%%%%%%%%%%%%%%%%%%%%%%%%%%%%

With the discovery of the Higgs boson at the LHC, one of the key questions was whether
it is elementary or composite. It used to be thought that a composite Higgs boson would
normally have a mass comparable to the scale of compositeness, but this can be
reduced if it is a pseudo-Nambu-Goldstone boson whose mass is protected by some approximate symmetry,
possibly compatible with the measured Higgs mass $\sim 125$~GeV.
The compositeness possibility may be probed using a phenomenological
Lagrangian ${\cal L}$ with free parameters to describe $H$ interactions, that may be
constrained using $H$ production and decay data.
In view of the consistency of the Standard Model relation $\rho \equiv m_W/m_Z \cos \theta_W = 1$ with data,
one may assume a custodial symmetry in this phenomenological Lagrangian: SU(2)$\times$SU(2) $\to$ SU(2). 
In this case, one may parametrize as follows the leading-order terms in ${\cal L}$:
\begin{eqnarray}
{\cal L} & = & \frac{v^2}{4} {\rm Tr} D_\mu \Sigma^\dagger D^\mu \Sigma \left( 1 + 2 a \frac{H}{v} + b \frac{H^2}{v^2} + \dots \right) \nonumber \\
& - & {\bar \psi}^i_L \Sigma \left(1 + c \frac{H}{v} + \dots \right) \nonumber \\
& + & \frac{1}{2} \left(\partial_\mu H \right)^2 + \frac{1}{2} m_H^2 H^2 + d_3 \frac{1}{6} \left(\frac{3 m_H^2}{v} \right) H^3 + d_4 \frac{1}{24} \left(\frac{3 m_H^2}{v} \right) H^4 + \dots \, ,
\label{calL}
\end{eqnarray}
where
\begin{equation}
\Sigma \; \equiv \; {\rm exp} \left( i \frac{\sigma^a \pi^a}{v} \right) \, .
\label{Sigma}
\end{equation}
The free coefficients $a, b, c, d_3$ and $d_4$ are all normalized such that they are unity
in the SM, whereas composite models may give observable deviations from these values.

\begin{figure}[htb]
\centering
\includegraphics[height=2.3in]{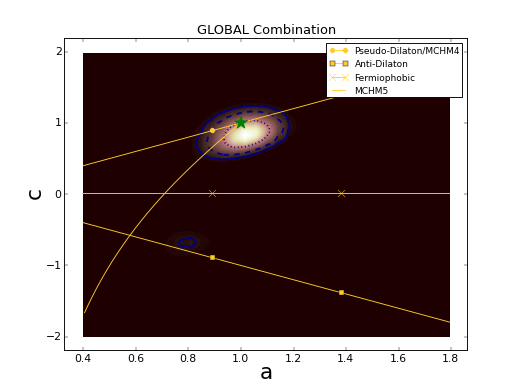}
\caption{\it A global fit to bosonic and fermionic $H$ couplings rescaled by factors $a$ and $c$, respectively,
indicating possible predictions of some composite models. The SM prediction $a = c = 1$ is shown as the green star~\protect\cite{EY3}.}
\label{fig:EY}
\end{figure}

Fig.~\ref{fig:EY} shows the result of one analysis~\cite{EY3},
that looked for possible rescalings of the $H$ couplings to bosons by a factor $a$
and to fermions by a factor $c$~\footnote{For a similar recent result from the CMS Collaboration, see~\cite{CMSmH}.
The ATLAS and CMS collaborations follow the Higgs Cross Section Working group in defining the
 quantities $\kappa_V \equiv a$ and $\kappa_f \equiv c$~\cite{HiggsxsWG}.}.
Clearly there is no sign of a significant deviation from the
SM prediction $a = c = 1$, and the specific composite models shown as yellow lines
in Fig.~\ref{fig:EY} are excluded unless (in some cases) their predictions can be adjusted to
be very similar those of the SM.

%%%%%%%%%%%%%%%%%%%%%%%%%%%%%%%%%%%%%%%%%%%%%%%%%%%%%%%%%%%%%%%%%%%%%%%%%
%%
%%   use this format to include an .eps figure into your paper
%%
%%%%%%%%%%%%%%%%%%%%%%%%%%%%%%%%%%%%%%%%%%%%%%%%%%%%%%%%%%%%%%%%%%%%%%%%%%%

\section{The SM as an Effective Field Theory}

In view of the continuing successes of the SM, now also in the Higgs sector, a common approach is
to regard it as an effective field theory valid at low energies $\lesssim 1$~TeV. The effects of higher-scale
physics may then be expressed, as a first approximation, via higher-dimensional operators constructed out of SM
fields, whose coefficients may be constrained by precision electroweak data, Higgs data
and triple-gauge couplings (TGCs). Table~\ref{tab:EWPToperators} lists the operators
entering electroweak precision tests (EWPTs) at LEP, together with $95\%$ CL bounds on their individual coefficients
when they are switched on one at a time, and also when marginalised in a simultaneous global fit~\cite{ESY4}.
For the first four coefficients we list the constraints from the leptonic observables alone,
while the constraints on the remaining coefficients also include hadronic observables. 

\begin{table}[h]
\begin{center}
\begin{tabular}
{|c | c | c | c | c | c |}
\hline
\multirow{2}{*}{Operator} & \multirow{2}{*}{Coefficient} & \multicolumn{2}{| c |}{LEP Constraints}  \\
 &  & Individual & Marginalized \\
\hline
${\mathcal O}_W=\frac{ig}{2}\left( H^\dagger  \sigma^a \leftrightarrow {D^\mu} H \right )D^\nu  W_{\mu \nu}^a$ & \multirow{2}{*}{$\frac{m_W^2}{\Lambda^2}(c_W+c_B)$} & \multirow{2}{*}{$(-0.00055, 0.0005)$}  & \multirow{2}{*}{$(-0.0033,0.0018)$}   \\
${\mathcal O}_B=\frac{ig'}{2}\left( H^\dagger  \leftrightarrow {D^\mu} H \right )\partial^\nu  B_{\mu \nu}$ &  &  &  \\
\hline
${\cal O}_T=\frac{1}{2}\left (H^\dagger {\leftrightarrow{D}_\mu} H\right)^2$ & $\frac{v^2}{\Lambda^2}c_T$ & $(0,0.001)$  & $(-0.0043, 0.0033)$  \\
\hline
$\mathcal{O}_{LL}^{(3)\, l}=( \bar L_L \sigma^a\gamma^\mu L_L)\, (\bar L_L \sigma^a\gamma_\mu L_L)$ & $\frac{v^2}{\Lambda^2}c^{(3)l}_{LL}$ & $(0,0.001)$  & $(-0.0013,0.00075)$  \\
\hline
${\mathcal O}_R^e =
(i H^\dagger {\leftrightarrow { D_\mu}} H)( \bar e_R\gamma^\mu e_R)$  & $\frac{v^2}{\Lambda^2}c^e_R$ & $(-0.0015,0.0005)$  & $(-0.0018,0.00025)$  \\
\hline
${\cal O}_{R}^u =
(i H^\dagger {\leftrightarrow { D_\mu}} H)( \bar u_R\gamma^\mu u_R)$ & $\frac{v^2}{\Lambda^2}c^u_R$ & $(-0.0035,0.005)$  & $(-0.011,0.011)$  \\
\hline 
${\cal O}_{R}^d =
(i H^\dagger {\leftrightarrow { D_\mu}} H)( \bar d_R\gamma^\mu d_R)$ & $\frac{v^2}{\Lambda^2}c^d_R$ & $(-0.0075,0.0035)$  & $(-0.042,0.0044)$ \\
\hline 
${\cal O}_{L}^{(3)\, q}=(i H^\dagger \sigma^a {\leftrightarrow { D_\mu}} H)( \bar Q_L\sigma^a\gamma^\mu Q_L)$ & $\frac{v^2}{\Lambda^2}c^{(3)q}_L$ & $(-0.0005,0.001)$  & $(-0.0044,0.0044)$ \\
\hline 
${\cal O}_{L}^q=(i H^\dagger {\leftrightarrow { D_\mu}} H)( \bar Q_L\gamma^\mu Q_L)$  & $\frac{v^2}{\Lambda^2}c^q_L$ & $(-0.0015,0.003)$  & $(-0.0019,0.0069)$  \\
\hline
\end{tabular}
\end{center}
\caption{\it Operators and coefficients contributing to LEP electroweak precision tests (EWPTs), 
includeng $95\%$ CL bounds when individual operators are switched on,
and also when they are marginalized in a simultaneous global fit~\protect\cite{ESY4}.}
\label{tab:EWPToperators}
\end{table}

\begin{figure}[h!]
\centering 
\includegraphics[scale=0.38]{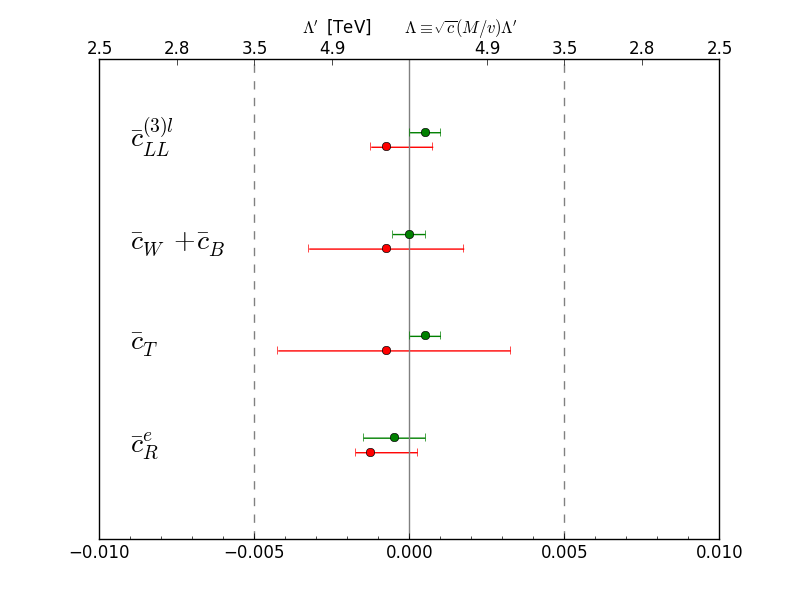}
\includegraphics[scale=0.38]{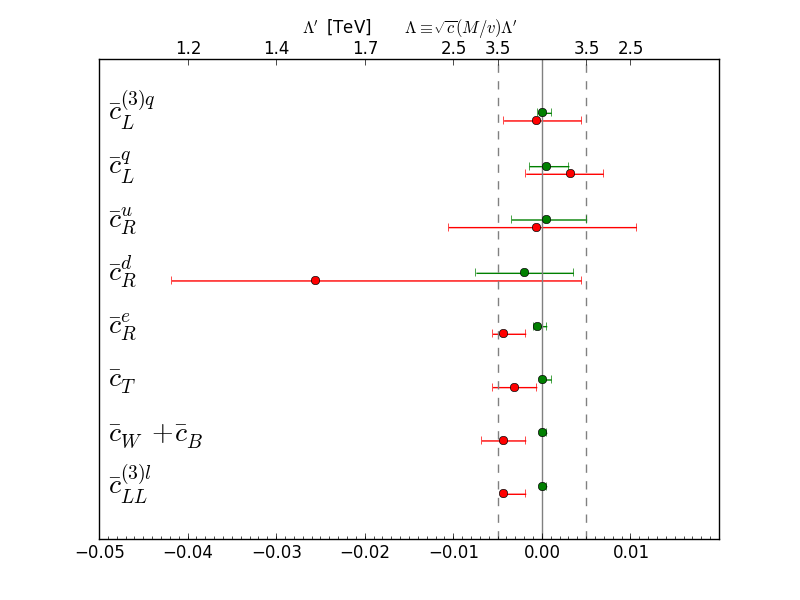}
\caption{\it The 95\% CL ranges found in analyses of the leptonic observables (left panel)
and including also the hadronic observables (right panel). In each case, the upper (green)
bars denote single-coefficient fits, and the lower (red) bars denote multi-coefficient fits.
The upper-axis should be read $\times \frac{m_W}{v}\sim 1/3$ for $\bar{c}_W + \bar{c}_B$.~\protect\cite{ESY4} }
\label{fig:EWPTsummary}
\end{figure}

Our results for fits to the coefficients $\bar{c}^{(3)l}_{LL}, \bar{c}_T, \bar{c}_W + \bar{c}_B$ and $\bar{c}^e_R$
that affect the leptonic observables $\{ \Gamma_Z, \sigma^0_\text{had}, R^0_e, R^0_\mu, R^0_\tau, A^{0,e}_\text{FB}, m_W \}$
are shown in the left panel of Fig.~\ref{fig:EWPTsummary}. 
The upper (green) bars show
the ranges for each of the coefficients varied individually, assuming that the other coefficients
vanish, and the lower (red) bars show the ranges for a global fit in which all the coefficients
are varied simultaneously. The legend at
the top of the left panel of Fig.~\ref{fig:EWPTsummary} translates the ranges of the coefficients
into ranges of sensitivity to a large mass scale $\Lambda$. We see that all the sensitivities are
in the multi-TeV range. 
The right panel of Fig.~\ref{fig:EWPTsummary} shows the effect of including the
hadronic observables $\{R^0_b, R^0_c, A^{0,b}_\text{FB}, A^{0,c}_\text{FB}, A_b, A_c\}$,
and the operator coefficients that contribute directly to them, namely $\bar{c}^q_L, \bar{c}^{(3)q}_L, \bar{c}^u_R$ and $\bar{c}^d_R$.     

\begin{figure}[h!]
\centering 
\includegraphics[scale=0.6]{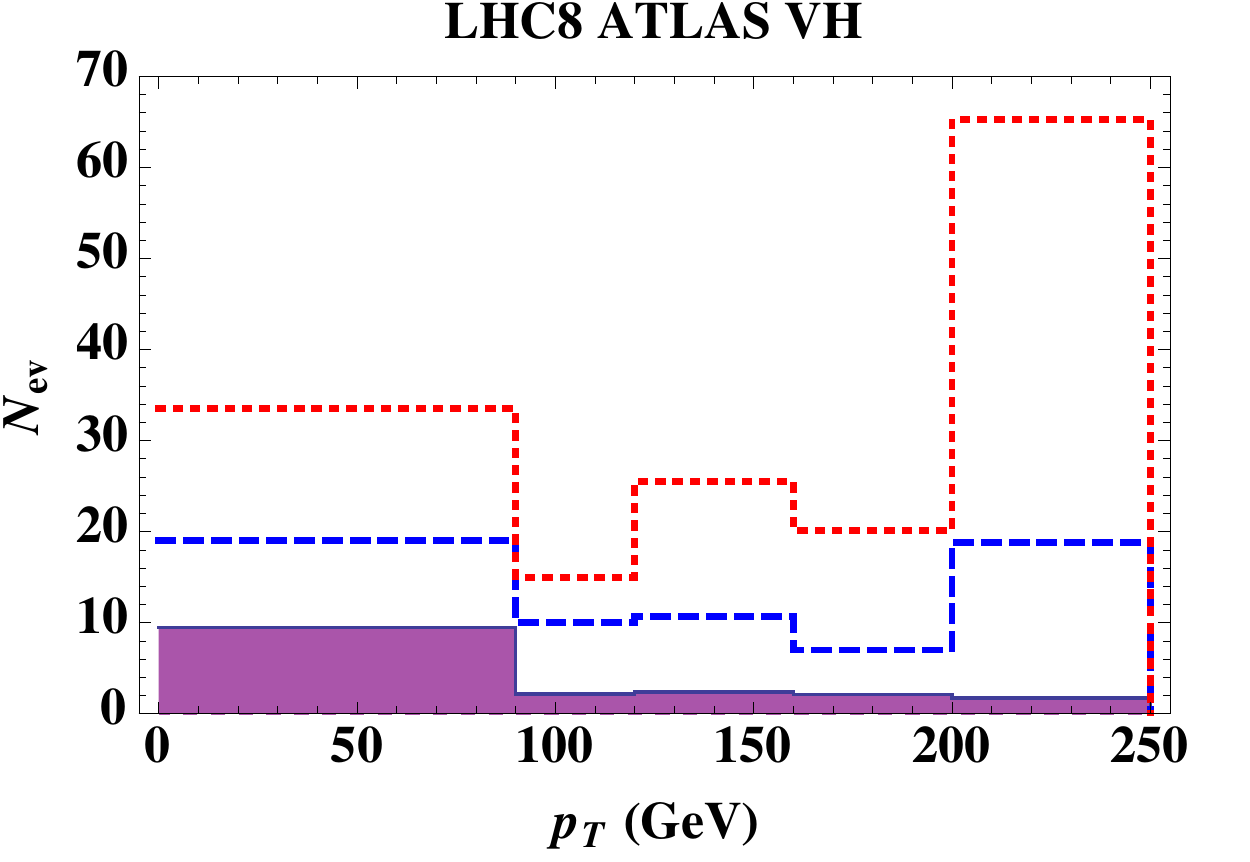}
\includegraphics[scale=0.35]{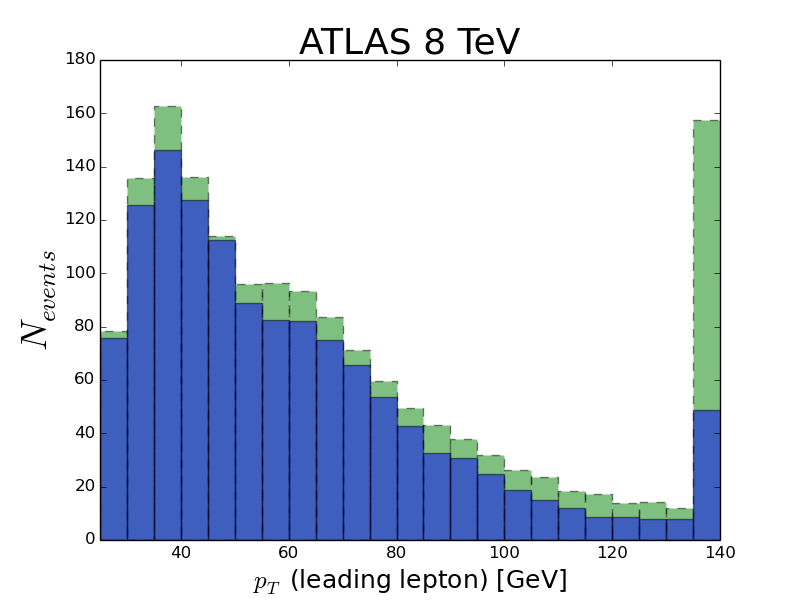}
\caption{\it Left panel: Simulation of the $p_T^V$ distribution in $(V \to 2 \ell) + (H \to {\bar b}b)$ events at the
LHC after implementing ATLAS cuts. 
The solid distribution is the SM expectation, and the red-dotted and blue-dashed lines 
correspond to the distributions with $\bar c_W$ =0.1 and 0.05, respectively~\protect\cite{ESYHV}.
Right panel: The same-flavour $p_T$ distribution of the leading lepton after the TGC analysis
cuts for ATLAS at 8 TeV. The Standard Model distribution is shown in blue with solid lines,
and the effect of $\bar{c}_{HW} = 0.1$ is superimposed in green with dashed lines.
In both cases the last (overflow) bin provides significant extra information compared to the overall normalisation~\protect\cite{ESY4}.}
\label{fig:distributions}
\end{figure}

The operators with the largest effects on Higgs physics and TGCs are listed in Table~\ref{tab:LHCoperators}~\footnote{EWPTs
constrain the operators in Table~1 so strongly that they are not important for Higgs physics and TGCs.}, 
together with $95\%$ CL bounds obtained when individual operators are switched on one at a time,
and also when they are marginalized in a simultaneous global fit~\cite{ESY4}. Important information is
provided by kinematic distributions~\cite{ESYHV}, as well as by total rates, as seen in Fig.~\ref{fig:distributions}.

\begin{table}[h]
\begin{center}
\begin{tabular}
{|c | c | c | c | c | c |}
\hline
\multirow{2}{*}{Operator} & \multirow{2}{*}{Coefficient} & \multicolumn{2}{| c |}{LHC Constraints}  \\
 &  & Individual & Marginalized \\
\hline
${\mathcal O}_W=\frac{ig}{2}\left( H^\dagger  \sigma^a \leftrightarrow {D^\mu} H \right )D^\nu  W_{\mu \nu}^a$ & \multirow{2}{*}{$\frac{m_W^2}{\Lambda^2}(c_W - c_B)$} & \multirow{2}{*}{$(-0.022,0.004)$}  & \multirow{2}{*}{$(-0.035,0.005)$}   \\
${\mathcal O}_B=\frac{ig'}{2}\left( H^\dagger  \leftrightarrow {D^\mu} H \right )\partial^\nu  B_{\mu \nu}$ &  &  &  \\
\hline
${\mathcal O}_{HW}=i g(D^\mu H)^\dagger\sigma^a(D^\nu H)W^a_{\mu\nu}$ & $\frac{m_W^2}{\Lambda^2}c_{HW}$ & $(-0.042,0.008)$ & $(-0.035,0.015)$  \\
\hline
${\mathcal O}_{HB}=i g^\prime(D^\mu H)^\dagger(D^\nu H)B_{\mu\nu}$ & $\frac{m_W^2}{\Lambda^2}c_{HB}$ & $(-0.053,0.044)$ & $(-0.045,0.075)$ \\
\hline
${\mathcal O}_{3W}= \frac{1}{3!} g\epsilon_{abc}W^{a\, \nu}_{\mu}W^{b}_{\nu\rho}W^{c\, \rho\mu}$ & $\frac{m_W^2}{\Lambda^2}c_{3W}$ & $(-0.083,0.045)$ & $(-0.083,0.045)$  \\
\hline
${\mathcal O}_{g}=g_s^2 |H|^2 G_{\mu\nu}^A G^{A\mu\nu}$ & $\frac{m_W^2}{\Lambda^2}c_{g}$ & $(0,3.0)\times 10^{-5}$ & $(-3.2,1.1)\times 10^{-4}$  \\
\hline
${\mathcal O}_{\gamma}={g}^{\prime 2} |H|^2 B_{\mu\nu}B^{\mu\nu}$ & $\frac{m_W^2}{\Lambda^2}c_{\gamma}$ & $(-4.0,2.3)\times 10^{-4}$ & $(-11,2.2)\times 10^{-4}$  \\
\hline
\end{tabular}
\end{center}
\caption{\it List of operators entering in LHC Higgs and TGC physics, 
together with $95\%$ CL bounds when individual coefficients are switched on one at a time, and 
marginalized in a simultaneous fit~\protect\cite{ESY4}.}
\label{tab:LHCoperators}
\end{table}

\begin{figure}[h!]
\centering
\includegraphics[scale=0.5]{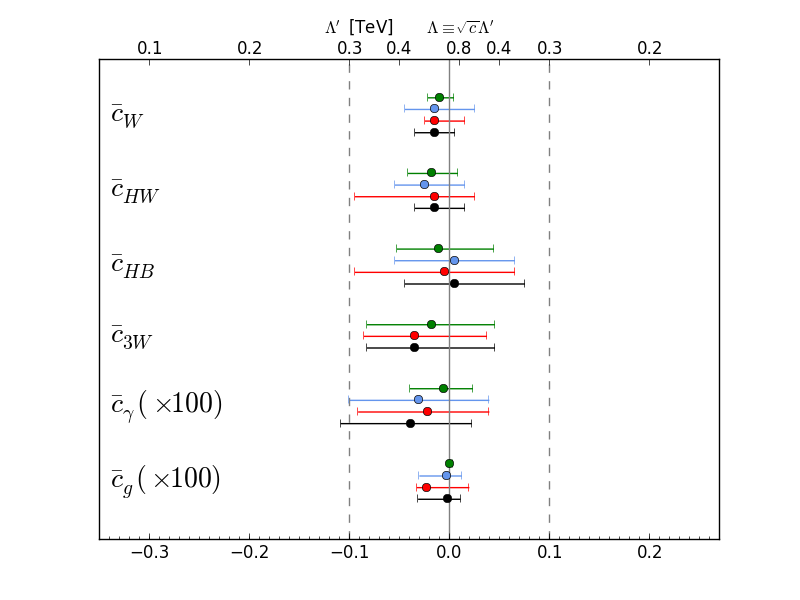}
\caption{\it The 95\% CL constraints obtained for single-coefficient fits (green bars),
and the marginalised 95\% ranges for the
LHC signal-strength data combined with the kinematic distributions for associated $H + V$ production
measured by ATLAS and D0 (blue bars), combined with the LHC TGC data (red lines), and the global combination with
both the associated production and TGC data (black bars). Note that $\bar{c}_{\gamma,g}$ are shown $\times 100$ for which the upper axis should therefore be read $\times 10$~\protect\cite{ESY4}.}
\label{fig:fitsummary}
\end{figure}

The results of a global fit to the Higgs data (including associated production kinematics)
and LHC TGC measurements are summarised in Fig.~\ref{fig:fitsummary}~\cite{ESY4}.
The individual 95\% CL constraints obtained by switching 
one operator on at a time are shown as green bars. The other lines are the marginalised 95\% ranges obtained using just the
LHC signal-strength data in combination with the kinematic distributions for associated $H + V$ production
measured by ATLAS and D0 (blue bars), in combination with the LHC TGC data (red lines), and in combination with
both the associated production and TGC data (black bars). We see that the LHC TGC constraints
are the most important for $\bar{c}_{W}$ and $\bar{c}_{3W}$, whereas the Higgs constraints are more
important for $\bar{c}_{HW}$, $\bar{c}_{HB}$ and $\bar{c}_{g}$.
Numerical results for the 95\% CL ranges for these coefficients are shown alongside the operator definitions in 
Table~\ref{tab:LHCoperators}.

\section{The SM is not enough!}

Albert Michelson infamously asserted in 1894 that {\it ``The more important
fundamental laws and facts of physical science have all been discovered"}, just 
before the discoveries of radioactivity and the electron. Likewise, Lord Kelvin asserted in 1900 that
{\it ``There is nothing new to be discovered in physics now,
all that remains is more and more precise measurement"}, just before Einstein
postulated the photon and proposed special relativity. Even after the discovery of a (the?)
Higgs boson, there are many reasons to expect physics beyond the SM, as I now discuss.

Inspired by James Bond~\cite{Bond},
here are 007 of them. 1) The measured values of $m_t$ and $m_H$ suggest that
the electroweak vacuum is {\it probably} unstable, unless some BSM physics intervenes.
2) The SM cannot provide the dark matter required by astrophysics and cosmology.
3) Additional CP violation beyond the CKM model is required to explain the origin
of the matter in the Universe.
4) The small sizes of the neutrino masses seem to require BSM physics.
5) New physics at the TeV scale is needed for the hierarchy of mass scales to seem more natural.
6) BSM physics is required for cosmological inflation, notably
because in the SM the effective Higgs potential would seem to become negative at high scales.
7) A consistent quantum theory of gravity would certainly require going (far) beyond the SM.
Some of these issues are discussed in the following.

\subsection{The Instability of the Electroweak Vacuum}

The effective electroweak potential of the SM resembles a Mexican hat,
invariant under the SM SU(2)$\times$U(1) symmetry. The origin is unstable, and is surrounded
by a valley where $\langle H \rangle \equiv v = 246$~GeV, the electroweak vacuum.
At larger Higgs field values, the brim of the hat rises, at least for a while.
However, calculations in the SM show that, for the measured values of $m_t$ and $m_H$
renormalization of the Higgs self-coupling by the top quark overwhelms that by the Higgs itself,
turning the hat brim down at large field values. Thus, in the SM the present electroweak vacuum is
unstable, in principle, with quantum tunnelling though the brim generating
collapse into an anti-de-Sitter 'Big Crunch'.

\begin{figure}[htb]
\centering
\includegraphics[height=2.1in]{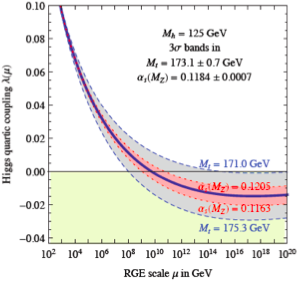}
\includegraphics[height=2.1in]{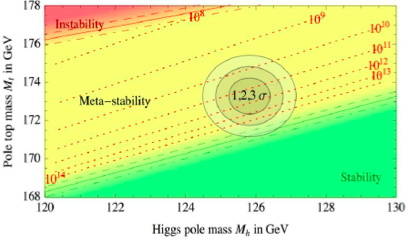}
\caption{\it Left panel: Within the SM, normalisation by the top quark appears to drive the Higgs self-coupling $\lambda < 0$.
Right panel: The regions of vacuum stability, metastability and instability in the $(m_H, m_t)$ plane.
Both panels are from~\protect\cite{Buttazzo}.}
\label{fig:Buttazzo}
\end{figure}

According to the SM calculations~\cite{Buttazzo} shown in the left panel of Fig.~\ref{fig:Buttazzo},
the brim turns down at a Higgs scale $\Lambda$:
\begin{equation}
\log_{10} \left( \frac{\Lambda}{{\rm GeV}} \right) \; = \; 11.3 + 1.0 \left(\frac{m_H}{{\rm GeV}} - 125.66 \right)
- 1.2 \left( \frac{m_t}{{\rm GeV}} - 173.10 \right) + 0.4 \left(\frac{\alpha_s(M_Z) - 0.1184}{0.0007} \right) \, .
\label{Buttazzo}
\end{equation}
Substituting the world average values of $m_t$, $m_H$ and $\alpha_s (M_Z)$, this formula yields
\begin{equation}
\Lambda \; = \; 10^{10.5 \pm 1.1}~{\rm GeV} \, .
\label{Lambda}
\end{equation}
This calculation is very sensitive to $m_t$, as seen in the right panel of Fig.~\ref{fig:Buttazzo}.
The D0 Collaboration has recently reported a new, higher, value of $m_t$~\cite{D0mt}, which would tend to
destabilise further the electroweak vacuum, whereas the CMS Collaboration has
reported lower values of $m_t$ from new analyses~\cite{CMSmt} that would tend to make
the vacuum more stable. A more accurate experimental measurement of $m_t$ would
help us understand the fate of the Universe within the SM, but this
experimental effort must be matched by improved understanding of the relationship
between the parameter $m_t$ in the SM Lagrangian and the effective mass parameter
appearing in the Monte Carlos used by experiments~\cite{Moch}.

%%%%%%%%%%%%%%%%%%%%%%%%%%%%%%%%%%%%%%%%%%%%%%%%%%%%%%%%%%%%%%%%%%%%%%%%%
%%
%%   use this format to include an .eps figure into your paper
%%
%%%%%%%%%%%%%%%%%%%%%%%%%%%%%%%%%%%%%%%%%%%%%%%%%%%%%%%%%%%%%%%%%%%%%%%%%%%

According to these calculations, the lifetime of the electroweak vacuum
would (probably) be much longer than the age of the Universe, but
this does not mean that one can simply ignore the problem.
We think that the Universe once had a very
high energy density, e.g., during an inflationary epoch~\cite{CMB}, and quantum and thermal
fluctuations at that time would have tended to populate
the anti-de-Sitter `Big Crunch' region of the effective potential~\cite{Oops}. Of course, we
might have struck lucky, living in a non-anti-de-Sitter region and thereby surviving - though this
would require anthropic special pleading~\footnote{It has
also been pointed out recently~\cite{BGM} that small black holes would catalyse rapid vacuum decay.}. The
problem could be avoided with suitable higher-dimensional
terms in the effective potential~\cite{Sher}, or other examples of new physics beyond the SM.
One example is supersymmetry~\cite{ER}, which would have
prevented the turn-down of the brim of the Mexican hat.

\subsection{Supersymmetry}

There are many longstanding reasons for loving supersymmetry (SUSY), such as 
alleviating the fine-tuning aspect of the hierarchy problem, providing
a natural candidate for the cold dark matter, facilitating grand unification
and playing an essential r\^ole in string theory. Run~1 of the LHC has added
three more reasons to this list, namely the mass of the Higgs boson, which is
within the range predicted by supersymmetry~\cite{SUSYmH,FH}, the fact that the couplings resemble
those of the SM Higgs boson, as expected in simple realisations of the minimal
supersymmetric extension of the SM (MSSM)~\cite{EHOW},
and the stabilisation of the electroweak vacuum, as mentioned at the end of
the previous Section. How could Nature resist SUSY's manifold charms?

However, despite our ardent love for SUSY, so far she has kept out of sight. Direct searches for SUSY at the LHC have
drawn blanks so far. This is also the case for searches for the scattering of dark matter particles,
indirect searches in flavour physics, etc.. How can we interpret these searches, and where may SUSY
be hiding? Unfortunately, we know that SUSY must be a broken symmetry, but we do not know how:
there are no signposts in superspace! It is often assumed that there is a residual discrete R-symmetry
that guarantees the stability of the lightest supersymmetric particle (LSP), which is the
dark matter candidate mentioned above. Beyond that, it is often assumed (without much
theoretical justification) that the SUSY-breaking sparticle masses are universal at some high
renormalisation scale, usually the GUT scale. The simplest model in which all the SUSY-breaking
contributions $m_0$ to the squark, slepton and Higgs masses are equal at the GUT scale, and
the SU(3), SU(2) and U(1) gauging masses $m_{1/2}$ are also universal, is called
the constrained MSSM (CMSSM). One may also consider models in which the SUSY-breaking
contributions to the masses of the two Higgs doublets of the MSSM are equal but different from
those of the squarks and leptons (the NUHM1), or where they are also different from each other (the NUHM2).

\begin{figure}[htb]
\centering
\includegraphics[height=2.3in]{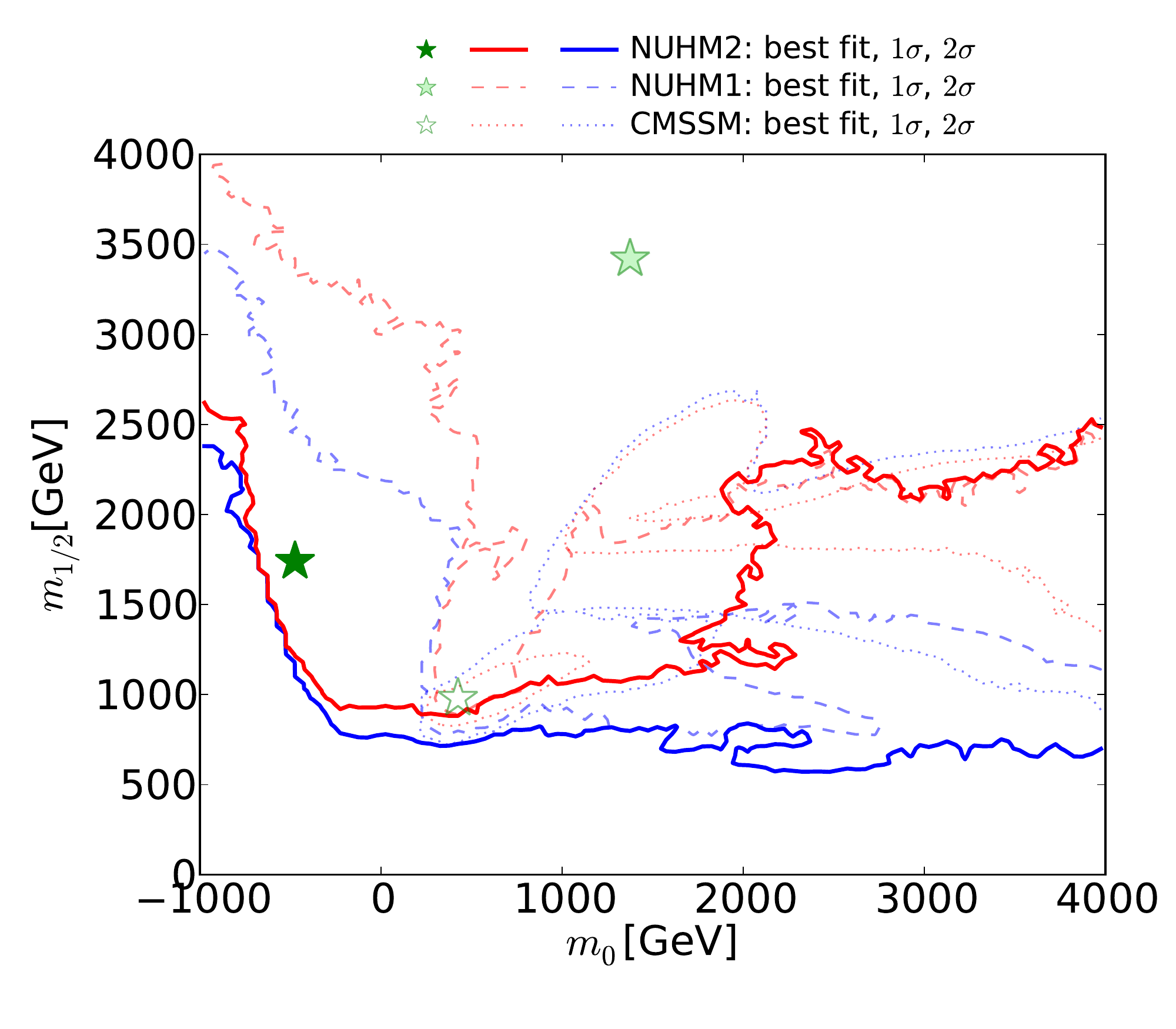}
\includegraphics[height=2.2in]{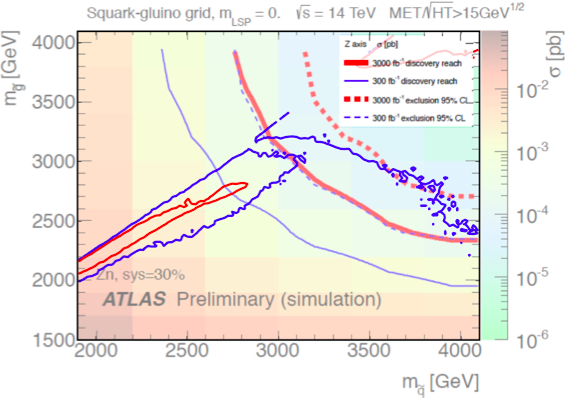}
\caption{\it Left panel: The 68\% CL (red) and 95\% CL contours (blue) in the $(m_0, m_{1/2})$ panes
 for the CMSSM (dotted lines), NUHM1 (dashed lines) and NUHM2 (solid lines)~\protect\cite{MC10}.
 Right panel: The reach of ATLAS in the $(m_{\tilde q}, m_{\tilde g})$ plane for exclusion and discovery
 with 300 and 3000/fb of integrated LHC luminosity at high energy~\protect\cite{ATLASfuture},
 compared with the 68\% CL (red) and 95\% regions in the CMSSM.}
\label{fig:MC10}
\end{figure}

\begin{figure*}[htb!]
%%%%%%%%%%%%%%%%%%%%%%%%%%%%%%%
%\vspace{-1cm}
\resizebox{7.5cm}{!}{\includegraphics{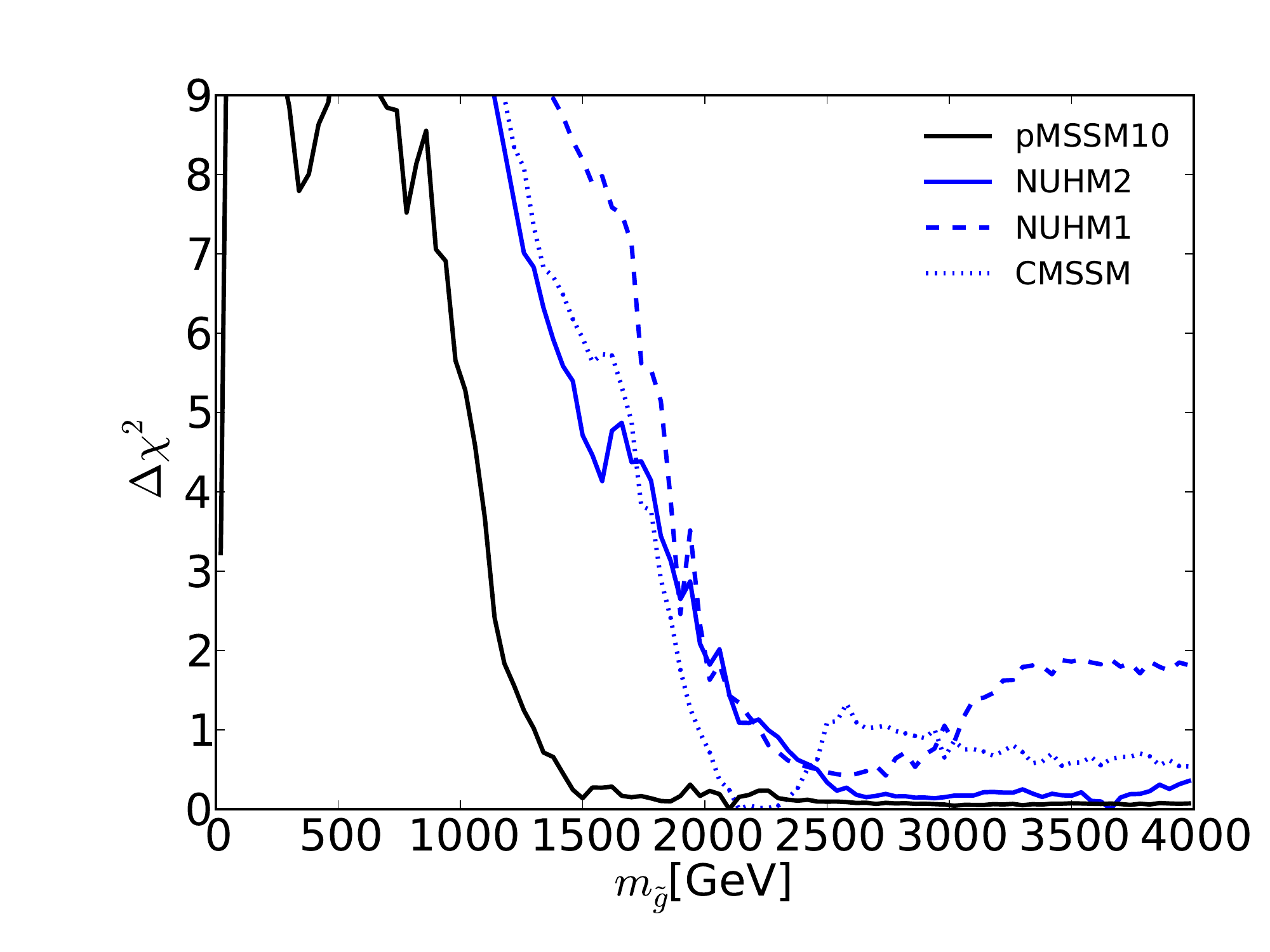}}
\resizebox{7.5cm}{!}{\includegraphics{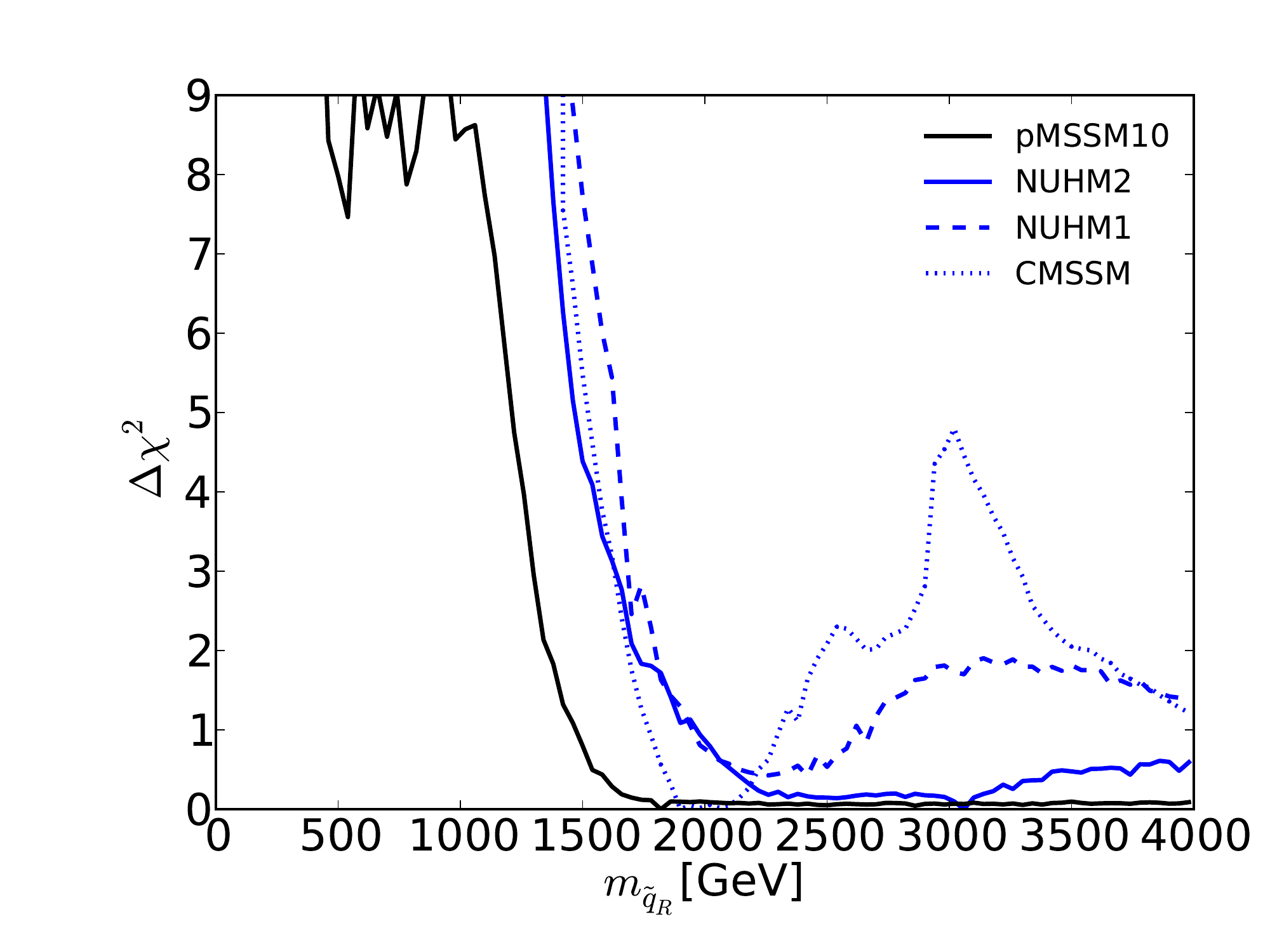}}\\
%%%%%%%%%%%%%%%%%%%%%%%%%%%%%%%
\vspace{-1cm}
\caption{\it The one-dimensional profile likelihood functions
for the gluino mass (left panel) and the first- and second-generation squark masses (right panel). 
In each panel the solid black line is for the pMSSM, the solid blue line for the NUHM2,
the dashed blue line for the NUHM1 and the dotted blue line for the
CMSSM~\protect\cite{MC11}.}
\label{fig:MC11}
\end{figure*}

The results of global fits to the CMSSM, NUHM1 and NUHM2,
combining all experimental and phenomenological constraints, and requiring that the relic supersymmetric particle density be
within the cosmological range, are shown projected on the $(m_0, m_{1/2})$ plane
in the left panel of Fig.~\ref{fig:MC10}~\cite{MC9,MC10}. In these models the 95\% CL
lower limits on the squark and gluino masses are $\sim 1.5$~GeV,
as seen in Fig.~\ref{fig:MC11}. The right panel of Fig.~\ref{fig:MC10} displays the
corresponding $(m_{\tilde q}, m_{\tilde g})$ plane, showing prospective exclusion and discovery reaches of the LHC
in future runs with 300 and 3000/fb of luminosity at high energy~\cite{ATLASfuture}. Superposed on this plane are the 68
and 95\% CL contours found in the global fit to the CMSSM. As already seen in the left panel of
Fig.~\ref{fig:MC10}, there are two distinct regions, the lower-mass one being favoured by the
disagreement between experiment and the SM prediction of $g_\mu - 2$.
We see that future runs of the LHC could detect squarks and gluinos if Nature is described by
supersymmetry with parameters in this lower-mass region of the CMSSM. 

%%%%%%%%%%%%%%%%%%%%%%%%%%%%%%%%%%%%%%%%%%%%%%%%%%%%%%%%%%%%%%%%%%%%%%%%%
%%
%%   use this format to include an .eps figure into your paper
%%
%%%%%%%%%%%%%%%%%%%%%%%%%%%%%%%%%%%%%%%%%%%%%%%%%%%%%%%%%%%%%%%%%%%%%%%%%%%

%%%%%%%%%%%%%%%%%%%%%% F I G U R E %%%%%%%%%%%%%%%%%%%%%%%%%%%%%%%%%%%
%%%%%%%%%%%%%%%%%%%%%% F I G U R E %%%%%%%%%%%%%%%%%%%%%%%%%%%%%%%%%%%

One may, instead, consider the phenomenological MSSM (pMSSM) in which no universality is assumed. In this case,
the the lower limits on the gluino and squark masses are reduced, compared with the CMSSM, NUHM1
and NUHM2, as seen in Fig.~\ref{fig:MC11}, enhancing the prospects for discovering SUSY in LHC Run~2~\cite{MC11}.
The left panel of Fig.~\ref{fig:MC11} displays the one-dimensional profile likelihood function
for the mass of the gluino in the pMSSM, and the right panel is the corresponding plot for
the first- and second-generation squarks.

\begin{figure}[htb]
\centering
\includegraphics[height=2.0in]{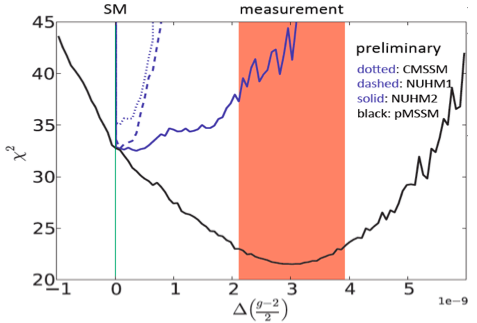}
\includegraphics[height=2.1in]{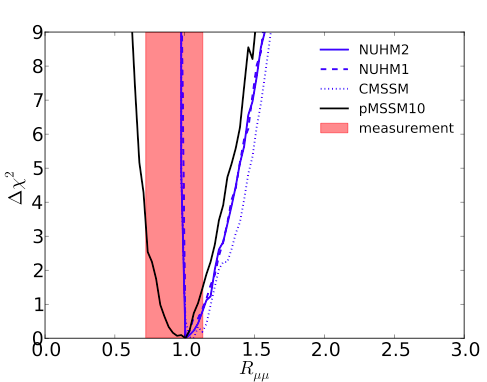}
\caption{\it The one-dimensional $\chi^2$ likelihood function for $g_\mu - 2$ (left panel) and $B_s \to \mu^+ \mu^-$
(right panel) in the CMSSM, NUHM1, NUHM2 and pMSSM~\protect\cite{MC11}. The vertical shaded bands
represent the 68\% CL ranges of $g_\mu - 2$ and $B_s \to \mu^+ \mu^-$.}
\label{fig:g-2}
\end{figure}

The pMSSM offers anew the possibility that supersymmetry could explain the discrepancy between the
SM calculation of $g_\mu - 2$ and the experimental measurement. The LHC constraints on
the CMSSM, NUHM1 and NUHM2 all predict values of the $g_\mu - 2$ that are very similar to the unsuccessful SM prediction,
as seen in the left panel of Fig.~\ref{fig:g-2}, whereas the experimental
measurement pMSSM could be accommodated within the pMSSM~\cite{MC11}.
Fortunately, there are plans for two new experiments to measure $g_\mu - 2$~\cite{futureg-2},
and other low-energy $e^+ e^-$ experiments will help refine the SM predictions, clarifying the discrepancy between
the SM and experiment. Turning to $B_s \to \mu^+ \mu^-$,
as seen in the right panel of Fig.~\ref{fig:g-2}, all of the CMSSM, NUHM1, NUHM2 and pMSSM
predict branching ratios very similar to the SM.

%%%%%%%%%%%%%%%%%%%%%%%%%%%%%%%%%%%%%%%%%%%%%%%%%%%%%%%%%%%%%%%%%%%%%%%%%
%%
%%   use this format to include an .eps figure into your paper
%%
%%%%%%%%%%%%%%%%%%%%%%%%%%%%%%%%%%%%%%%%%%%%%%%%%%%%%%%%%%%%%%%%%%%%%%%%%%%

\section{CP-Violating MSSM Scenarios}

\begin{figure}[!t]
 \begin{center}
\includegraphics[width=7.5cm]{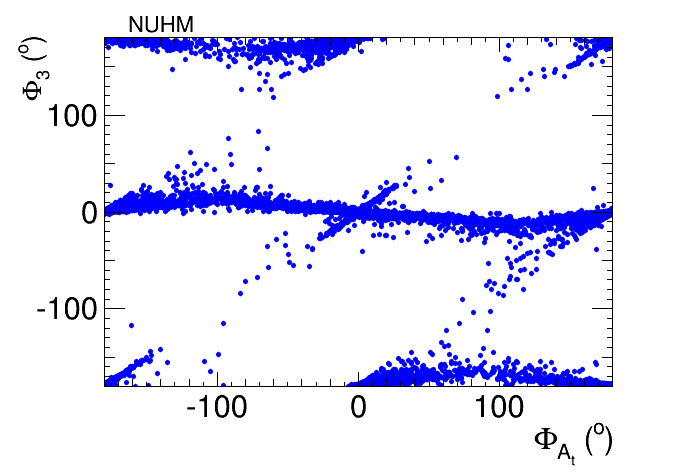}\includegraphics[width=7.5cm]{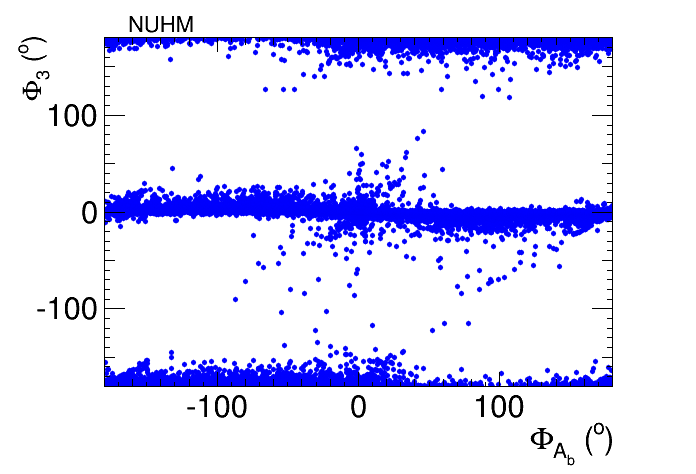}
\caption{\it Correlations of the phase of $M_3$ with that of $A_t$ (left panel) and
with that of $A_b$ (right panel) imposed by the EDM constraints in the NUHM2 scenario~\protect\cite{AEGM}.}
\label{fig:CPV}
 \end{center}
\end{figure}

In the CMSSM and related models, one can introduce 6 CP-violating phases, even if one assumes minimal flavour violation:
there are 3 phases in the gaugino masses $M_{1,2,3}$, and 3 more in the third-generation trilinear SUSY-breaking
couplings $A_{t,b,\tau}$~\cite{MCPMFV}. There are 4 important constraints on these phases coming from upper limits on the
electric dipole moments (EDMs) of the neutron, thorium monoxide, thallium and mercury. These 4 constraints in the 6-dimensional
CP-violating parameter space leave a 2-dimensional blind subspace where combinations of the phases may be large,
as seen in the case of the NUHM2 in Fig.~\ref{fig:CPV}~\cite{AEGM}. 
In the left panel we see the correlation these constraints impose between
the phases of $M_3$ and $A_t$, and in the right panel the correlation between the phases of $M_3$ and $A_b$.
In both cases we see diagonal features including
populations of points with large phases: the EDM constraints do not require all the CP-violating
phases to be small simultaneously.

\begin{figure}[!t]
 \begin{center}
  \includegraphics[width=7.5cm]{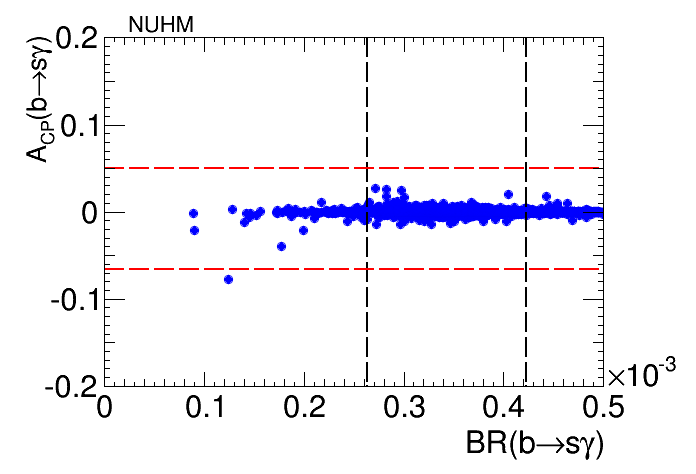} 
  \includegraphics[width=7.5cm]{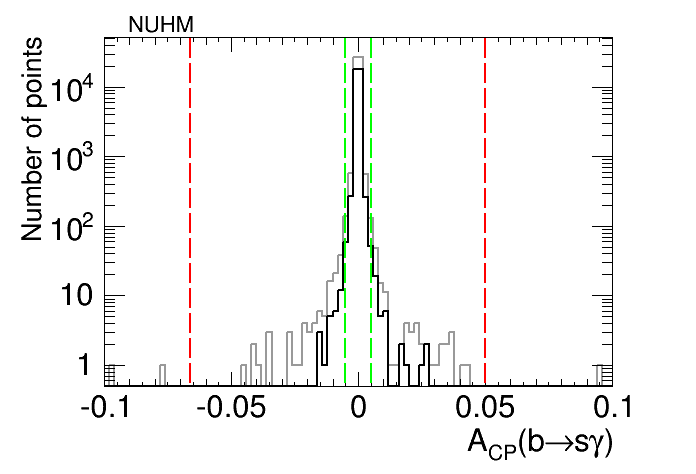}
\caption{\it Left panel: Scatter plot of the branching ratio for $b\to s\gamma$ decay versus its
CP-violating asymmetry, $A_{CP}$, in the NUHM2 scenario. Right panel: Histogram
of $A_{CP}$ in the NUHM2, imposing only the Higgs mass and EDM cuts (grey: full sample, black:
points satisfying the EDM constraints). The vertical red
dashed lines represent the present experimental limits, and the vertical green dashed lines
represent the prospective future improvement in the sensitivity to $A_{CP}$ 
by a factor of ten~\protect\cite{AEGM}.\label{fig:nuhm-ACPbsg}}
 \end{center}
\end{figure}

We have explored what values of the CP-violating
asymmetry in $b \to s \gamma$ decays, $A_{CP}$, are compatible with the EDM constraints.
We find that values of $A_{CP} \lesssim 2$\% can be found in the NUHM2 for values
of the $b \to s \gamma$ branching ratio lying within the experimentally-allowed range,
as shown in the left panel of Fig.~\ref{fig:nuhm-ACPbsg}, and the right panel of
this Figure displays a histogram of the NUHM2 results (grey: full sample, black:
points satisfying the EDM constraints). The vertical red dashed lines are
the present experimental constraints on $A_{CP}$~\cite{Agashe:2014kda}.
According to a study of the prospective Belle II sensitivity~\cite{Aushev:2010bq},
it should be possible to improve the present experimental sensitivity by a factor of ten,
as shown by the vertical green dashed lines, and we see that there are some CP-violating
NUHM2 models that could be explored with such an improvement.
We conclude that the EDM constraints allow an
observable value of $A_{CP}$ within the NUHM2, and the same is true in the pMSSM.

\section{Dark Matter Searches}

\begin{figure}[htb]
\centering
\includegraphics[height=2.3in]{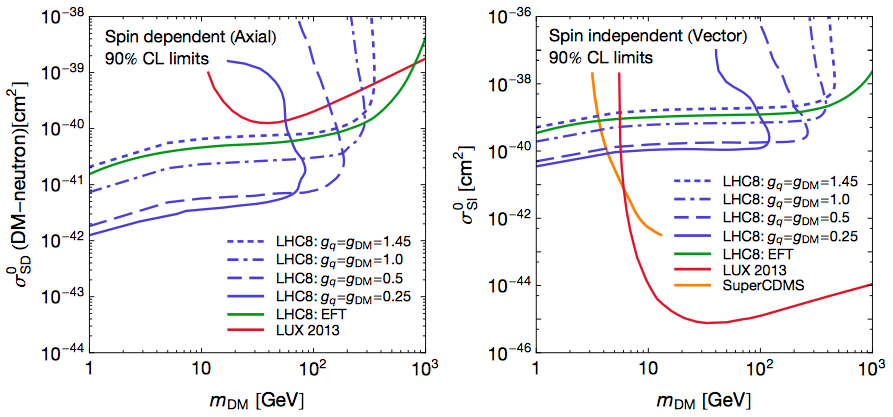}
\caption{\it A comparison of the current 90\% CL direct search limits from LUX and SuperCDMS
(red and orange lines, respectively), the monojet limits in simple models (blue lines) and the limits in
an effective field theory framework (green line) in the cross section vs $m_{DM}$ plane used by the direct detection community.
The left and right panels show, respectively, the limits on the spin-dependent and spin-independent cross sections
appropriate for axial- vector and vector mediators~\protect\cite{ICDM}.}
\label{fig:DMcomparison}
\end{figure}

As already commented, if a supersymmetric model conserves R-parity it provides a natural
candidate for a cold dark matter particle. The same is true in some other TeV-scale extensions
of the SM, such as some extra-dimensional models with K-parity and little Higgs models with T-parity.
The subject of searches for TeV-scale dark matter particles is vast, and there is no time or space here to discuss
it in detail. However, I take the opportunity to display in Fig.~\ref{fig:DMcomparison} a recent
comparison between the current reaches of the LHC via monojet searches and direct
astrophysical searches for the scattering of generic TeV-scale dark matter particles,
for the cases of spin-dependent (axial) couplings (left panel) and spin-independent (vector) couplings
(right panel)~\cite{ICDM}. We see that in the former case the LHC monojet searches generally have greater sensitivity
than the direct searches, except for dark matter particle masses $\gtrsim 1$~TeV where the LHC
runs out of phase space. On the other hand, direct searches for spin-independent interactions are
stronger than the LHC searches for masses $\gtrsim 4$~GeV. Supersymmetry models generally
favour a relatively large mass for the dark matter particle.

%%%%%%%%%%%%%%%%%%%%%%%%%%%%%%%%%%%%%%%%%%%%%%%%%%%%%%%%%%%%%%%%%%%%%%%%%
%%
%%   use this format to include an .eps figure into your paper
%%
%%%%%%%%%%%%%%%%%%%%%%%%%%%%%%%%%%%%%%%%%%%%%%%%%%%%%%%%%%%%%%%%%%%%%%%%%%%

\section{Possible Future Colliders}

\begin{figure}[htb]
\centering
\includegraphics[height=3in]{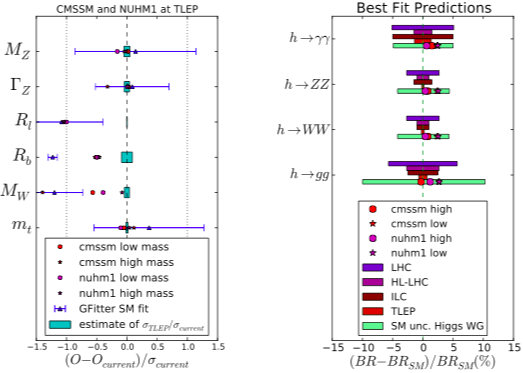}
\caption{\it Comparison of the present precisions in measurements of various $Z$ properties (left panel) and Higgs couplings
(right panel), together with the prospective precisions of possible measurements at future colliders,
the current theoretical uncertainties, and the deviations from the SM predictions found at the best-fit points in various SUSY models. From~\protect\cite{TLEP}.}
\label{fig:TLEP}
\end{figure}

In conclusion, here are a few words about possible future colliders. There has long been interest in building
a high-energy linear $e^+ e^-$ collider such as the ILC ($E_{CM} \lesssim 1$~TeV) or CLIC ($E_{CM} \lesssim 3$~TeV),
and there is now increasing interest in Europe and China in a possible large circular tunnel that could accommodate either an
$e^+ e^-$ collider with $E_{CM} \lesssim 350$~GeV and/or a $pp$ collider with $E_{CM} \lesssim 100$~TeV~\cite{FCC}. A circular
$e^+ e^-$ collider could also provide unparalleled accuracy in measuring the properties of the $Z$ and Higgs bosons,
in particular, as seen in Fig.~\ref{fig:TLEP}~\cite{TLEP}. In principle, these could be able to distinguish between the predictions of
the SM and various fits in the CMSSM, NUHM1 and NUHM2, as shown. However, this will require considerable
efforts to reduce correspondingly the present theoretical uncertainties, shown by the shaded green bars.

%%%%%%%%%%%%%%%%%%%%%%%%%%%%%%%%%%%%%%%%%%%%%%%%%%%%%%%%%%%%%%%%%%%%%%%%%
%%
%%   use this format to include an .eps figure into your paper
%%
%%%%%%%%%%%%%%%%%%%%%%%%%%%%%%%%%%%%%%%%%%%%%%%%%%%%%%%%%%%%%%%%%%%%%%%%%%%

\begin{figure}[!t]
 \begin{center}
  \includegraphics[width=7.5cm]{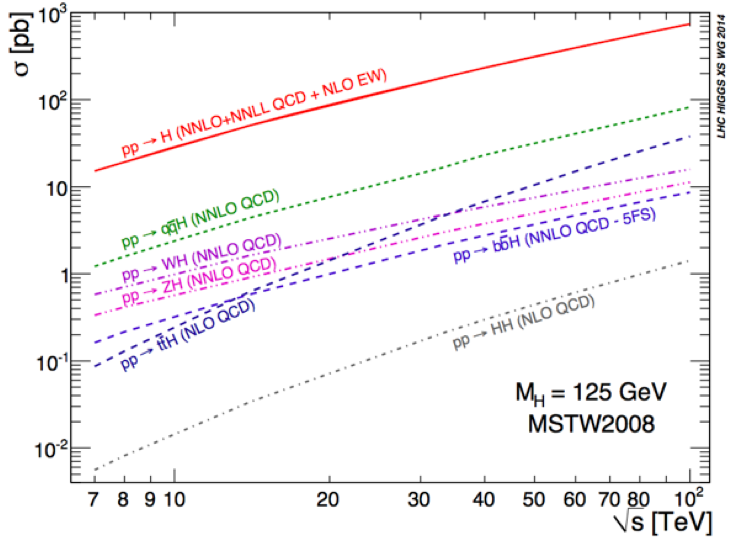} 
  \includegraphics[width=7cm]{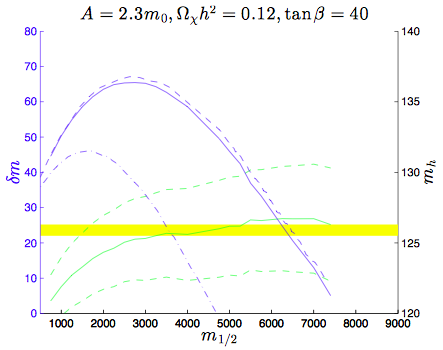}
\caption{\it Left panel: Cross sections for various Higgs production processes at $pp$ colliders
as functions of the centre-of-mass energy~\protect\cite{FCC-hh}. Right panel: One of the possibilities for a relatively
heavy supersymmetric dark matter particle weighing $\sim 0.4 m_{1/2} \lesssim 3$~TeV. The
vertical axis is the mass difference between the dark matter particle and the next-to-lightest
supersymmetric particle, in this case the lighter stop squark. The solid and upper dashed blue lines
correspond to the current central and +1$\sigma$ values of the dark matter density, the horizontal
yellow band represents the experimental value of the Higgs mass, and the green
lines represent the central value and $\pm$1$\sigma$ uncertainties in
theoretical calculations of the Higgs mass~\protect\cite{EOZ}.}
\label{fig:FCC-hh}
 \end{center}
\end{figure}

A future high-energy $pp$ collider would also offer the possibility of producing very large numbers of
Higgs bosons, as seen in the left panel of Fig.~\ref{fig:FCC-hh}~\cite{FCC-hh}, and studies are underway to estimate
better the accuracies experiments could give in measuring Higgs couplings, including in particular the
elusive triple-Higgs coupling, which will be very difficult to measure at the LHC or in any but
a very high-energy $e^+ e^-$ collider. A high-energy $pp$ collider would also offer unique possibilities
to discover and/or measure the properties of supersymmetric particles. Even the lightest of these
could weigh several TeV, as seen in the right panel of Fig.~\ref{fig:FCC-hh}~\cite{EOZ}. In the example shown,
only a $pp$ collider with $E_{CM} \sim 100$~TeV would be capable of exploring the full range of
particle masses compatible with SUSY providing dark matter weighing $\lesssim 3$~TeV (solid
and upper dashed blue lines), and the green lines show that for all this range calculations of the
Higgs mass are compatible with the experimental value (represented by the yellow band),
considering the theoretical uncertainties represented by the green lines.

The physics cases for future large circular colliders are still to be developed. Certainly there
will be a strong bedrock of high-precision Higgs and other SM measurements to test possible
scenarios for physics beyond the SM. As for direct searches for new physics, the search
for dark matter particles looks to provide the strongest case, but this still requires further
study. Needless to say, the physics landscape will look completely different when/if future
runs of the LHC at higher energy and luminosity provide some evidence for new physics beyond
the SM. Some of my supersymmetric friends are disappointed that SUSY has not shown up yet,
but the LHC saga has only just begun. It took 48 years for the Higgs boson to be discovered~\cite{Economist}
(the record time-lag so far), but
four-dimensional supersymmetric were first written down only 41 years ago~\cite{WZ}, so let us be patient
for a while longer!

\ack
I am grateful to Nick Mavromatos and Sarben Sarkar for their kind invitation to speak
at this fun conference. The work was supported in part by the London Centre for Terauniverse Studies
(LCTS), using funding from the European Research Council via the Advanced Investigator
Grant 267352 and from the UK STFC via the research grant ST/J002798/1.

\end{document}